\documentclass[twocolumn,aps,pra,showpacs]{revtex4}
\usepackage{epsfig}
\begin{document}
\title{Improved storage of coherent and squeezed states in imperfect ring cavity}
\author{Petr Marek and Radim Filip}
\affiliation{Department of Optics, Palack\' y University,\\
17. listopadu 50,  772~07 Olomouc, \\ Czech Republic}
\date{\today}
\begin{abstract}
We propose a method of an improving quality of a ring cavity which is imperfect due
to non-unit mirror reflectivity. The method is based on using squeezed states of light pulses 
illuminating the mirror and gradual homodyne detection of a 
radiation escaping from the cavity followed by single
displacement and single squeezing operation performed on the released
state. We discuss contribution of this method in process
of storing unknown coherent and known squeezed state and generation of squeezing in the optical ring cavities.
\end{abstract}
\pacs{03.65.Ud}
\maketitle

\section{Introduction}

Recently, quantum information processing and quantum communication experiments utilizing photons as information carrier 
(for a review \cite{book}) are performed without an utilizing advantage of quantum memories. 
The quantum memories can store a quantum state carrying information for a further processing 
as well as they can store an entanglement resource for an actual time quantum communication. 
However, the photons are relatively difficult to store and an implementation of a practical quantum memory 
for photons remains a challenging problem. Earlier proposals for a quantum memory 
were mainly dedicated to a storing quantum states of individual photons 
in a high-Q cavity \cite{cav}, in collective atomic excitations \cite{atom} or in a fiber loop \cite{memory}.
To enhance storing individual photon in the fiber loop, a linear-optical quantum computing circuit that runs 
an error-correction code \cite{qec} was proposed \cite{Gingrich03}. To protect a qubit against decoherence, 
the schemes based on the decoherence-free subspace \cite{dfs} were also proposed and
implemented for trapped ions \cite{dfsexp}. 

Quantum information processing with continuous variables (CVs)
based on a manipulation with Gaussian states of many-photon systems \cite{CV} 
represents an interesting alternative to the quantum information and communication protocols exploiting 
individual photons. In the CV quantum information protocols, the coherent states
are mainly utilized as the information carrier and the squeezed states
are basic resource for a production of the CV entangled states. To perform more complex and collective CV
quantum information protocols we would like to implement a quantum memory which is able to 
store unknown coherent state or known squeezed state for a long time. 
To store a continuous-variable information encoded as simultaneous amplitude and phase modulation of coherent state 
for a long time we can utilize a simple quantum memory device based on optical-fiber loop or a ring cavity. 
We lock the continuous-wave field into the high-fidelity fiber loop or ring cavity and release it when we need. 
Recently, a similar method has been proposed to store the polarization qubit in a 
long length fiber \cite{memory}. However, in the ring cavity an unavoidable 
losses at mirrors deteriorate information encoded in the coherent state and typically restricts the storage time.
Further, to produce a pronouncedly squeezed state for efficient CV quantum information processing, a
ring cavity containing a nonlinear crystal is frequently used in a continuous-wave 
optical parametric oscillator/amplifier (OPO/OPA) 
\cite{ring}. Here an unavoidable losses of the imperfect mirrors restricts a maximal 
value of the produced squeezing. A method of the squeezing generation in the ring cavity was used, for example, 
to produce entangled state for a teleportation of coherent states \cite{tele}.  
 
\begin{figure}
\centerline{\psfig{width=8.0cm,angle=0,file=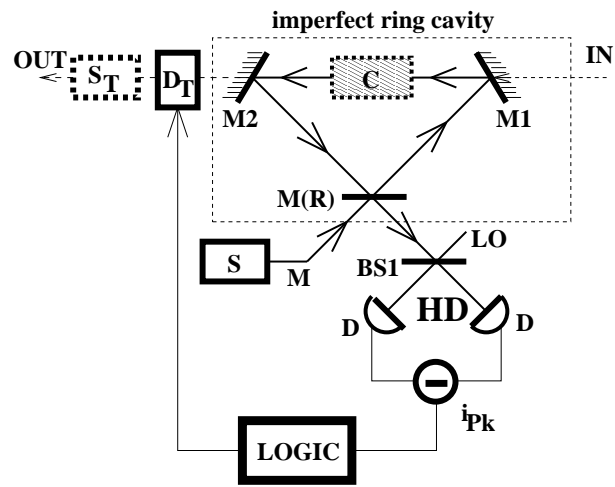}}
\caption{Setup for protection of the coherent, squeezed states and 
squeezed state generation: $C$ -- nonlinear crystal, $M1,M2$
-- high-quality mirrors, $M(R)$ -- imperfect mirror with the reflectivity $R$, $S$ -- source of the squeezed states, $LO$ --
local oscillator, $BS1$ -- 50:50 beam splitter, $HD$ -- homodyne measurement, $D$ -- detectors, $D_{T}$ -- displacement operation,
$S_{T}$ -- squeezing operation.}
\end{figure}

In this paper we propose a method how to
increase the quality of the ring cavity if we are able to inject an auxiliary squeezed light to the cavity 
through the imperfect mirror and detect a field leaving outside from the mirror by a homodynne detection in every 
cycle of the field in the ring cavity. 
According to the measured data, a simply joint feed-forward correction by a modulation of the final state 
can be performed after many cycles in the ring cavity and we can 
preserve an unknown coherent or known squeezed state for a long time as will be shown in Sec.~II. 
For the storing of the coherent state it is even necessary only to perform an 
additional squeezing operation to obtain a fidelity of the 
storing which is independent on unknown input amplitude. 
The presented method is based on CV quantum erasing procedure \cite{Filip03e} which allows
us to restore at least partially the input state after an interaction.
Thus a reflectivity of the mirror is in fact enhanced 
using the erasing procedure. It can also used to enhance maximal 
amount of the squeezing generated in OPO/OPA in which 
the ring cavity is filled by a nonlinear crystal as will be demonstrated in Sec.~III. 
Thus we could consequently stimulate an increase of fidelity of CV quantum information protocols.
This method is different to the error correcting codes since it works only with information which 
leaves from the memory unit. 

\section{Protection of CV state}

In this Section, we demonstrate usefulness of quantum erasing to
achieve an enhancement of a time of the storage of unknown
coherent state or known squeezed state. We consider an empty
 imperfect ring cavity (without the crystal $C$) depicted in
Fig.~1 which consists of two mirrors $M1,M2$ with almost unit reflectivity 
at a frequency of stored field. The third imperfect mirror $M(R)$ has a less reflectivity
$R$ than $M1,M2$ however still $R>0.99$ as is typical in this kind of experiments
experiments. The rest of the losses and imperfections are assumed
to be negligible in our analysis.

To demonstrate a protection by the erasing effect, the standard
setup of imperfect cavity (in the box) is completed by the
generator $S$ of squeezed states, balanced homodynne detection $HD$,
displacement correction $D_{T}$ and squeezing correction $S_{T}$.
The erasing procedure can be performed as follows. In every
round-trip of the quantum field in cavity, a state squeezed in the
quadrature variable $X_{M}$ is mixed with a state of an internal
cavity mode and the output state leaving from the mirror $BS$ is
detected by homodyne measurement producing the current $i_{P}$
proportional to a measured quadrature of the field. During many
cycles in the cavity, the currents $i_{P1},i_{P2},\ldots,i_{PN}$
corresponding to every cycle are registered in a computer memory.
After $N$ cycles corresponding to storage time, a quantum state
leaving the cavity by a reducing the reflectivity of the mirror $M2$ can be corrected by
a total displacement $D_T$ calculated from the measured values
$i_{P1},i_{P2},\ldots,i_{PN}$, known reflectivity $R$ and total
number $N$ of the round trips. In addition, we can convert the
corrected state to another having the same mean values as the
input state by total squeezing operation $S_T$, depending on the
reflectivity $R$ and number $N$ of the cycles.

Now, let us look at this procedure in detail. Using Heisenberg
picture representing each mode of light by a pair of the conjugate
quadrature operators $X$ and $P$ satisfying the commutation
relations $[X,P]=i$, the k-th pass through the cavity mirror
$M(R)$ can be represented by transformation relations
\begin{eqnarray}\label{Prot 1}
X_k = RX_{k-1} + TX_{Mk}, & P_{k} = RP_{k-1} + T P_{Mk} \nonumber \\
X_{Mk}' = TX_{k-1} -RX_{Mk}, & P_{Mk}' = TP_{k-1} - RP_{Mk},
\end{eqnarray}
where $X_k$, $P_k$ denote the quadrature variables of the signal
mode after $k$-th round and $X_{Mk}$, $P_{Mk}$ stand for the
quadrature variable of the meter mode used to inject a squeezed
state into the mirror. Detecting a field from the mirror by
the homodyne detection, the operator $P_{Mk}'$ collapses on
a real number $i_{Pk}$. Using relation (\ref{Prot 1}) we can
straightforwardly derive an evolution of the quadrature operators
\begin{eqnarray}\label{Prot 2}
    X_N = R^N X_{in} + T \sum_{k=1}^{N} R^{k-1}  X_{Mk}, \nonumber
    \\
    P_N = R^{-N} P_{in} - T \sum_{k=1}^{N}R^{-k} i_{Pk}
\end{eqnarray}
after $N$ round trips in the cavity. Here $X_{in}$ and $P_{in}$
describe the initial quadrature operators of the signal mode which
we are trying to protect. Now to suppress the decoherence effect
in the signal mode after the series of the $N$ passages through
the mirror, we can use all measured values $i_{P1},\ldots,i_{PN}$
and implement on signal mode the total displacement operation
\begin{equation}\label{Prot 3}
    X_N' = X_N, \quad P_N' = P_N + T \sum_{k=1}^{N}R^{-k}
    i_{Pk},
\end{equation}
followed by additional squeezing operation
\begin{equation}\label{Prot 4}
    X_{out} = \frac{X_N'}{R^N}, \quad P_{out} = R^N P_N'
\end{equation}
to achieve an universal character of the protection. Universality
means that the mean values of both complementary variables are
preserved. Thus the resulting universal transformation
\begin{equation}\label{Prot 5}
    X_{out} = X_{in} + \frac{T}{R^N} \sum_{k=1}^{N} R^{k-1}
    X_{Mk}, \quad P_{out} = P_{in}
\end{equation}
of the quadrature operators is obtained. Through our activities
any unknown input state was fully restored in the momentum.
Further, the mean value of both the coordinate and momentum are
unchanged if the injected states have vanishing mean values of the
quadratures. However the variance of the quadrature $X_{out}$ will
differ from the variance $X_{in}$. If we consider the injected
states to be independent and having the same variance
$\sigma_{XM}$ , we can write the variance of $X_{out}$ in the
following form
\begin{equation}\label{Prot 6}
   \sigma_{Xout}^2 = \sigma_{Xin}^2 + (R^{-2N} -1) \sigma_{XM}^2.
\end{equation}
From this follows that we can reduce the noise in the quadrature
$X_{out}$ as well if we use a state squeezed in the coordinates
$X_{Mk}$ and thus achieve almost perfect protection of an unknown
coherent or known squeezed state. The reason is that the 
mixing the signal with appropriately squeezed states on the mirror $M(R)$ approaches
the ideal quantum non-demolition measurement of the certain field
quadrature, which can be reversed perfectly by the erasing
procedure \cite{Filip03e}. Note that an amount of fluctuations in
the complementary variables $P_{Mk}$ does not influence the
proposed protection procedure and only the squeezing of single
quadrature is relevant.

For a comparison, without any measurements performed, the resulting variances would look as
\begin{eqnarray}\label{Prot 6}
\sigma_{Xout}^2 = R^{2N} \sigma_{Xin}^2 + (1-R^{2N})\sigma_{XM}^2, \nonumber \\
\sigma_{Pout}^2 = R^{2N} \sigma_{Pin}^2 + (1-R^{2N})\sigma_{PM}^2,
\end{eqnarray}
if the protected state has mean values equal to zero. However, if
this was not true, we would have to perform phase-insensitive amplification to
achieve universality and the variances would be
\begin{eqnarray}\label{Prot 7}
\sigma_{Xout}^2 =  \sigma_{Xin}^2 + (R^{-2N} -1) \sigma_{XM}^2 +
(R^{-2N}-1) \sigma_{Xz}^2 ,\nonumber \\
\sigma_{Pout}^2 =  \sigma_{Pin}^2 + (R^{-2N} -1) \sigma_{PM}^2 +
(R^{-2N} -1) \sigma_{Pz}^2,
\end{eqnarray}
with $X_z$, $P_z$ being the operators of ancillary mode of the
amplifier. Qualitatively, the protection method reduces noise
completely in the quadrature $P_{out}$ and also partially in the quadrature $X_{out}$ as can be seen
from Eq.~(\ref{Prot 6},\ref{Prot 7})

\begin{figure} \label{FigCoh}
\centerline{\psfig{width=8.0cm,angle=0,file=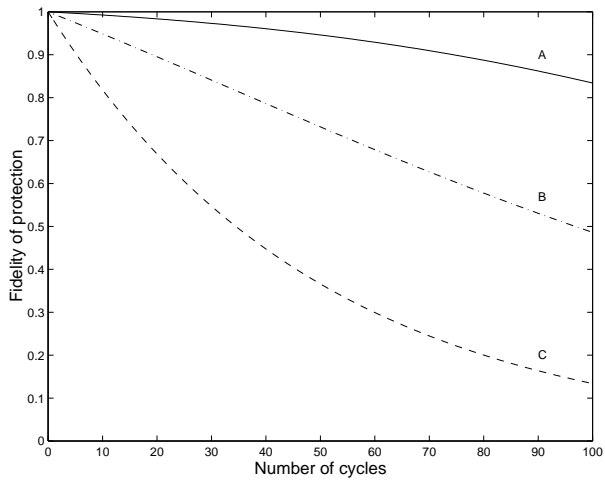}}
\caption{Fidelity of a coherent state in the protected ring cavity: $R= 0.99$,
variance of squeezed quadrature of the meter modes $\sigma_M^2 =
0.5\exp(-2)$.}
\end{figure}
\begin{figure}\label{FigSq}
\centerline{\psfig{width=8.0cm,angle=0,file=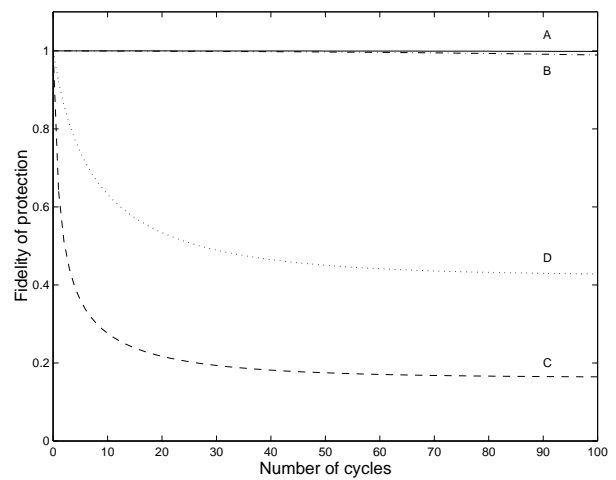}}
\caption{Fidelity of a squeezed vacuum state in the protected ring cavity: $R=
0.99$, the variance of squeezed quadrature of the meter modes $\sigma_M^2
= 0.5\exp(-2)$, signal mode squeezed in the quadrature $P$ with
the variance $\sigma_{Pin}^2 = 0.5\exp(-5)$.}
\end{figure}

To demonstrate an usefulness of the protection quantitatively
we fix the reflexivity $R$ of the mirror $M(R)$ and the squeezing in the
auxiliary modes $Mk$ since we will compare different strategies
with the same resources. In CV quantum computing there are
important two basic types of quantum states, coherent and squeezed
vacuum state, and our aim was to show how the protection we
proposed increases efficiency of storage of these states. We
consider and compare four different strategies: (A) protection
with vacuum squeezed in $Xk$ injected into the modes $Mk$ and
homodynne detections, (B) protection with unsqueezed vacuum
injected in the modes $Mk$ and homodynne detections, (C)
unprotected storing and (D) only injecting vacuum squeezed in $Pk$
in into the modes $Mk$ without the homodynne detections.

To compare the different strategies we use the fidelity of protection
\begin{equation}\label{Prot 8}
    F = \left[ (\sigma_{Xin}^2 + \sigma_{Xout}^2) (\sigma_{Pin}^2 + \sigma_{Pout}^2)
    \right]^{-1/2},
\end{equation}
which is the overlap between initial and resulting Gaussian
state. Due to universality of storage only the variances of 
the initial and resulting state must to be calculate to infer the fidelity. 
For unknown initial coherent state, the fidelity of protection strongly depends on squeezing of meter
mode. In Fig.~\ref{FigCoh} we use for illustration a feasible
squeezed state generated with the variance $\sigma_M^2 =
0.5\exp(-2)$ corresponding to 3dB squeezing injected to the
imperfect mirror with $R=0.99$. It is evident that using erasing
with the squeezed state is the best strategy for a long time
preservation of unknown coherent state. We can see a qualitative
change from an exponential decoherence in the case $C$ to the
non-exponential one in the case $A$. The strategy of squeezed
vacuum injection (D) is counterproductive since any unbalancing
noise in the complementary quadratures at output results in
formidable decrease of fidelity.

On the other hand, squeezed states are asymmetrical in the variances 
so that we can greatly benefit from asymmetry of our method, if we know
the orientation of the squeezed state which could be protected. In our method,
a noise in a quadrature is suppressed at the cost of blurring
the other one. Since our correction completely preserves the
squeezed quadrature and the noise added into the conjugate one is
diminutive compared to noise already present, the fidelity is
almost unity. Also the squeezing of the mode $Mk$ is not of great
importance for almost perfect reflectivity $R$, as can be seen in
Fig.~(\ref{FigSq}). A larger distinction between the strategies
(A) and (B) arises for a large number $N$ of the rounds in the
cavity. Both other strategies (C,D) are worse than (A,B). Here the
fidelity strongly depends on the proportion of variances of
squeezed quadratures of the signal and meter modes. If we consider
the input state is squeezed substantially more then the meter
modes, what is reasonable demand, we never find the fidelity for
strategy (D) comparable with the protected one. It turned out that
for both unknown coherent and known squeezed state protection is
always the best strategy.

\section{ Protection of squeezing generation}

Because of a strong squeezing producing a large entanglement is required for the CV
quantum information protocols, the optical cavity often used to
form OPO/OPA so that the down-converted fields pass the nonlinear
medium numerous times. For the single pass case, the interaction
in continuous-wave regime is weak and requested gain is obtained
after sufficient number of rounds in the cavity, thus equivalently
lengthening the interaction distance. As an example we assume ring
cavity filled by a nonlinear crystal exhibiting (degenerate or
non-degenerate) down-conversion process. It consists of two mirrors $M1,M2$ with almost unit 
reflectivity at a frequency of down-converted beams and almost unit transitivity for the pump
beam. This setup operating as a continuous-wave frequency degenerate but polarization
non-degenerate subtreshold OPO with collinear phase-matched type-II
down-conversion in KTP was frequently used in the many experiments
\cite{ring}. It produces a squeezed state in two linear
polarization modes which can be simply converted to single mode
squeezing by $\lambda/2$-wave plate along a direction $\pi/4$
relative to these polarizations. The nonlinear crystal is pumped
by a pulse from the ring laser with intra-cavity frequency
doubling. For simplicity, a pumping part of the experimental setup
was omitted in Fig.~1.

Next we consider that a nonlinear crystal $C$ producing squeezed
light pumped by an intensive laser pulse is inserted into the
cavity to obtain a source of a sufficiently squeezed state. Here
the imperfection of the cavity mirror decreases the efficiency of
the squeezing generation, possibly hindering the cumulation of
effect at all. We can use the previous method of protection to
effectively reduce a losses in  mirror $M(R)$ and thus enhance an
efficiency of the squeezing generation.

To protect the operation we apply a slight modification of
above described procedure. The ring cavity containing
nonlinear medium performing a weak squeezing of the signal field
during each trip in the cavity can be described by the following transformation relations
\begin{eqnarray}\label{SQ 1}
X_k = RGX_{k-1} + TX_{Mk}, & P_{k} = \frac{R}{G}P_{k-1} + T P_{Mk} \nonumber \\
X_{Mk}' = GTX_{k-1} -RX_{Mk}, & P_{Mk}' = \frac{T}{G}P_{k-1} -
RP_{Mk},
\end{eqnarray}
for the quadrature operators after the $k$-th cycle. Here $G$ is a
gain of squeezing per single cycle in the cavity which is typically small. 
In an analogy with the previous case, after the displacement operation
\begin{equation}\label{SQ 2}
    X_N' = X_N, \quad P_N' = P_N + \frac{T}{R} \sum_{k=1}^{N}(RG)^{1-k}
    i_{Pk},
\end{equation}
 we find the resulting state have the variances
\begin{eqnarray}\label{SQ 2}
    \sigma_{Xout}^2 = \frac{1}{2}(RG)^{2N} + T^2 \frac{1 -
    (RG)^{2N}}{1-(R G)^2}\sigma_{XM}^2, \nonumber
    \\
    \sigma_{Pout}^2 =\frac{1}{2}(RG)^{-2N}.
\end{eqnarray}
if we consider unsqueezed vacuum to be the initial state. Note, in contrast
to protection of the state,  we have not performed any additional
squeezing $S_{T}$ of the signal mode after it leaved the cavity.

Again we will compare few different strategies, so for evaluation
is needed to find the variances of signal field quadrature
operators 
\begin{eqnarray}\label{SQ 5}
 \sigma_{Xout}^2 = \frac{1}{2}(R G)^{2N} +  T^2
    \frac{1 - (R G)^{2N}}{1 - (R G)^2}\sigma_{XM}^2,
    \\
    \sigma_{Pout}^2 = \frac{1}{2} (\frac{R}{G})^2 +  T^2 \frac{1 -
   (\frac{R}{G})^{2N}}{1 - (\frac{R}{G})^2}\sigma_{PM}^2
\end{eqnarray}
if no active protection was performed.
Recall now, we are trying to protect the squeezing operation. That
is, if our initial state was vacuum then our target state should
be a minimal uncertainty squeezed state. There are two criteria of
a quality that we have taken into our account. First one is overall
fidelity between the target and actually produced state, the
second one is an amount of achievable single quadrature squeezing.
Again, four different strategies (A,B,C,D) were considered as in
the previous Section.

\begin{figure}\label{FigTar}
\centerline{\psfig{width=8.0cm,angle=0,file=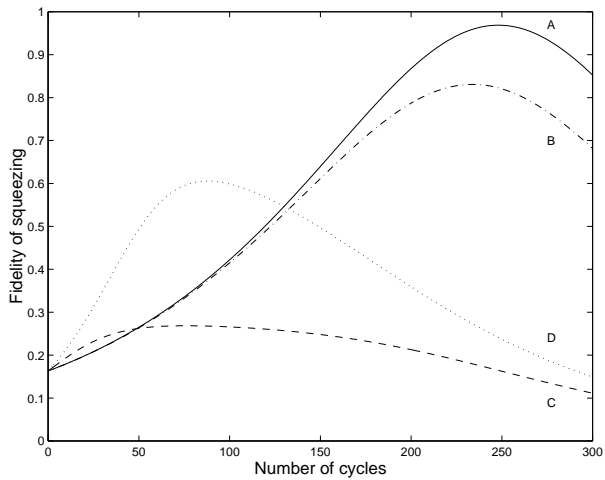}}
\caption{Fidelity of squeezed state produced in the protected ring cavity: $R= 0.99$, the gain of
squeezing in a single cycle $G=\exp(0.02)$, the variance of the squeezed quadrature of meter
modes $\sigma_M^2 = 0.5\exp(-2)$. A target mode is squeezed in
quadrature $P$ with variance $\sigma_{Ptarget}^2 = 0.5\exp(-5)$.}
\end{figure}

\begin{figure}\label{FigVP}
\centerline{\psfig{width=8.0cm,angle=0,file=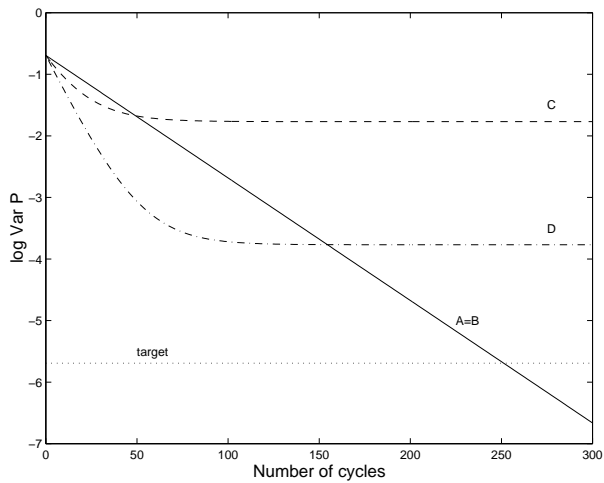}}
\caption{Logarithm of squeezed variance of state produced in the protected ring cavity: $R= 0.99$, the gain of
squeezing in a single cycle $G=\exp(0.02)$, the variance of squeezed quadrature of meter
modes $\sigma_M^2 = 0.5\exp(-2)$, A target mode is squeezed in
quadrature $P$ with variance $\sigma_{Ptarget}^2 = 0.5\exp(-5)$.}
\end{figure}


Consider a pure squeezed state as a target state and ask how is it
close to our prepared state. To quantify it we can count the
fidelity
\begin{equation}\label{B08}
    F = \left[ (\sigma_{Xout} + \sigma_{Xtarget})(\sigma_{Pout}
    +\sigma_{Ptarget})\right]^{-1/2}.
\end{equation}
between target and obtained state and compare it using the
strategies (A,B,C,D). The result is depicted in Fig.~\ref{FigTar}.
For a small number of cycles the strategy of squeezed vacuum injection (D)
looks better but only because meter mode squeezing is still
comparable or even better than the squeezing of signal field, so
interaction on the mirror actually improves the state in the
cavity. However, at some point the relevant quadrature cannot be
squeezed further due to the cavity losses and each cycle in cavity
only adds additional noise to the conjugate quadrature what
results in a fidelity decrease. If we apply the correction
procedure the squeezing buildup is slower, but inevitable. We can
represent squeezing protected by our method as nearly ideal
(squeezed quadrature ideal, antisqueezed quadrature slightly
disturbed), but with the gain $RG$. From this interpretation is
apparent we need reflexivity and gain choose so that $RG > 1$ to
obtain high fidelity results.

Next we can study how the evolution of noise in the quadrature
which we are trying to squeeze depend on the number $N$ of cycles.
The results can be understand from Fig.~\ref{FigVP}. The
logarithmic scale was chosen to clearly show what is happening.
First, for protected case (A,B), the variance of momentum is not
dependent on squeezing in modes $Mk$ and it has obviously no
theoretical lower bound. Here also is not necessary to use a
squeezed meter modes for a generation of the squeezing in a single
quadrature with no respect to the noise in the complementary one.
On the other hand, for the strategies (C,D) when the protection is
not performed the squeezed variances saturate at a point
\begin{equation}\label{bound}
    \lim_{N \to \infty} \sigma_{Pout} = G^2
    \frac{1-R^2}{G^2 - R^2}\sigma_{PM}
\end{equation}
determined by the proprieties of the cavity and the used squeezing
in the modes $Pk$. In Fig.~\ref{FigVP} we use in the case (D) the
same amount of squeezing $\sigma_MP = 0.5\exp(-2)$ as in the cases
(A,B) however in the complementary quadrature.

From this follows, if we are interested only in achieving
squeezing in one quadrature, what is often sufficient in the CV
quantum information protocols, such as optimal CV teleportation,
or protection of state depicted in previous section, it is not
necessary use a squeezing in the modes $Mk$ to protect the
squeezing. We need only perform an effective homodyne measurement
followed by a single displacement operation at the end.

\section{Conclusion}
We have shown how collective quantum erasure performed on single
mode in different times can be used to increase the quality of the
cavity and the efficiency of storing and squeezing generation
processes within. Our method formidably increases the fidelity of
storing of unknown coherent state, but due to its asymmetrical
nature allows almost perfect storage of known squeezed state. If
our protocol is applied to process of squeezing generation it
removes the lower bound for one quadrature squeezing attainable
and also allows production of pure squeezed states with high
fidelity.

\medskip
\noindent {\bf Acknowledgments}

The work was supported by project 202/03/D239 of
The Grant Agency of the Czech Republic, by the projects LN00A015
and CEZ: J14/98 of the Ministry of Education of Czech Republic.


\begin{thebibliography}{99}

\bibitem{book}
The Physics of Quantum Information, D. Bouwmeester, A. Ekert and A. Zeilinger (eds.), Springer-Verlag (2000). 

\bibitem{cav}
X. Maitre, E. Hagley et al., Phys. Rev. Lett. {\bf 79}, 769 (1997). 

\bibitem{atom}
D. F. Phillips et al.,
Phys. Rev. Lett. {\bf 86}, 783-786 (2001); J. M. Taylor et al., Phys. Rev. Lett. {\bf 90}, 206803 (2003); 
C. Schori et al., Phys. Rev. Lett. {\bf 89}, 057903 (2002). 
Ch. van der Wal, Science {\bf 301}, 196 (2003)

\bibitem{memory}
T.B. Pittman and J.D. Franson, Phys. Rev. A {\bf 66}, 062302 (2002);

\bibitem{Gingrich03}
R.M. Gingrich, quant-ph/0306098. 

\bibitem{qec}
P.W. Shor, Phys. Rev. A {\bf 52}, 2493 (1995); D. Gottesman, Phys. Rev.
A {\bf 54}, 1862 (1996); A. Ekert and C. Macchiavello, Phys. Rev. Lett.
{\bf 77}, 2585 (1996); A.R. Calderband {\em et al.}, Phys. Rev. Lett. {\bf 78}, 405
(1997).

\bibitem{dfs}
P. Zanardi, and M. Rasetti, Phys. Rev. Lett. {\bf 79}, 3306 (1997);
D.A. Lidar, I.L. Chuang, and K.B. Whaley, Phys. Rev. Lett. {\bf 81}, 2594
(1998); L.M. Duan, and G.C. Guo, Phys. Rev. A {\bf 57}, 737 (1998); 

\bibitem{dfsexp}
D. Kielpinski, et al., Science {\bf 291}, 1013 (2001);

\bibitem{CV}
S.L. Braunstein, Phys. Rev. Lett. {\bf 98}, 869 (1998); S.L. Braunstein and H.J. Kimble, 
S. Lloyd and S.L. Braunstein, Phys. Rev. Lett. {\bf 82}, 1784 (1999);
Phys. Rev. A {\bf 61}, 042302 (2000); S.F. Pereira et al., Phys. Rev. A {\bf 62}, 042311 (2000);  
Ch. Silberhorn, et al., Phys. Rev. Lett. {\bf 86}, 4267 (2001);
S.L. Braunstein et al., Phys. Rev. Lett. {\bf 86}, 4938 (2001); 
J. Fiur\' a\v sek, Phys. Rev. Lett. {\bf 86}, 4942 (2001);
X. Li et al., Phys. Rev. Lett. {\bf 88}, 047904 (2002); 
W.P. Bowen et al., Phys. Rev. Lett. {\bf 88}, 093601 (2002);
Ch. Silberhorn et al., Phys. Rev. Lett. {\bf 88}, 167902 (2002);  Ch. Silberhorn et al., 
Phys. Rev. Lett. {\bf 89}, 167901 (2002); 
F. Grosshans and P. Grangier and Phys. Rev. Lett. {\bf 88}, 057902 (2002)
W. P. Bowen et al., Phys. Rev. A {\bf 67}, 032302 (2003); W.P. Bowen et al., 
Phys. Rev. Lett. {\bf 90}, 043601 (2003); A. Dolinska et al., Phys. Rev. A {\bf 68}, 052308 (2003); 
F. Grosshans et al., Nature (London) {\bf 421}, 238 (2003);

\bibitem{ring}
L.-A. Wu et al., Phys. Rev. Lett. {\bf 57}, 691 (1986);
S.F. Pereira et al., Phys. Rev. A {\bf 38}, 4931 (1988); Z.Y. Ou, et al., Appl. Phys. B {\bf 55}, 
265 (1992); Z.Y. Ou et al., Phys. Rev. Lett. {\bf 68}, 3663 
(1992); Z.Y. Ou et al., Phys. Rev. Lett. {\bf 70}, 3239 (1992); S.F. Pereira, et al., 
Phys. Rev. Lett. {\bf 72}, 214 (1994); P.K. Lam, et al., J. Opt. B: Quantum Semiclass. Opt. {\bf 1}, 
469 (1999); B. Buchler, et al., Phys. Rev. A {\bf 65}, 011803(R) (2001); Y.J. Lu and Z.Y. Ou, Phys. Rev. A 
{\bf 62}, 033804 (2000);  

\bibitem{tele}
A. Furusawa, et al., Science {\bf 282}, 706 (1998); T.C. Zhang et al., quant-ph 0207076. 

\bibitem{Filip03e} 
R. Filip, Phys. Rev. A {\bf 67} 042111 (2003). 


\bibitem{feed}
P.K. Lam et al., J. Opt. B: Quantum Semiclass. Opt. {\bf 1}, 469 (1999); 
B. Buchler, et al., Phys. Rev. A {\bf 65}, 011803 (2001); U.L. Andersen, et al., 
J. Opt. B: Quantum Semiclass. Opt. {\bf 4}, S229 (2002).
  

\end{thebibliography}
\end{document}